# Automated Phase Mapping with AgileFD and its Application to Light Absorber Discovery in the V-Mn-Nb Oxide System

Santosh K. Suram,[a] Yexiang Xue,[b] Junwen Bai,[c] Ronan Le Bras,[b] Brendan Rappazzo,[b] Richard Bernstein,[b] Johan Bjorck,[b] Lan Zhou,[a] Robert B. van Dover,[d] Carla P. Gomes,[b,*] and John M. Gregoire[a,*]

[a] Joint Center for Artificial Photosynthesis, California Institute of Technology, Pasadena CA 91125, USA. E-mail: gregoire@caltech.edu

[b] Department of Computer Science, Cornell University, Ithaca, NY 14850, USA. E-mail: gomes@cs.cornell.edu

[c] Zhiyuan College, Shanghai Jiao Tong University, Shanghai, China

[d] Department of Materials Science and Engineering, Cornell University, Ithaca, NY, USA 14850

**Abstract:**

Rapid construction of phase diagrams is a central tenet of combinatorial materials science with accelerated materials discovery efforts often hampered by challenges in interpreting combinatorial x-ray diffraction datasets, which we address by developing AgileFD, an artificial intelligence algorithm that enables rapid phase mapping from a combinatorial library of x-ray diffraction patterns. AgileFD models alloying-based peak shifting through a novel expansion of convolutional nonnegative matrix factorization, which not only improves the identification of constituent phases but also maps their concentration and lattice parameter as a function of composition. By incorporating Gibbs' phase rule into the algorithm, physically meaningful phase maps are obtained with unsupervised operation, and more refined solutions are attained by injecting expert knowledge of the system. The algorithm is demonstrated through investigation of the V-Mn-Nb oxide system where decomposition of eight oxide phases, including two with substantial alloying, provides the first phase map for this pseudo-ternary system. This phase map enables interpretation of high-throughput band gap data, leading to the discovery of new solar light absorbers and the alloying-based tuning of the direct-allowed band-gap energy of $MnV_2O_6$. The open-source family of AgileFD algorithms can be implemented into a broad range of high throughput workflows to accelerate materials discovery.

**Introduction:**

Combinatorial materials science encompasses a suite of techniques for accelerated discovery of functional materials and often entails the establishment of relationships between crystal structure and materials properties.[1] As a result, high throughput phase mapping is an important tool for materials discovery, prompting the recent development of synchrotron-based x-ray diffraction (XRD) techniques to acquire $10^2$-$10^4$ XRD patterns over a composition spread library.[2] More generally, the accelerated characterization of phase diagrams in high-order composition spaces is a foundational problem in materials science given that many binary phase diagrams are known but few ternary or higher-order phase diagrams have been explored due to the requisite number of experiments. Emboldened by the progress in previous decades in automated XRD analysis for small molecules,[3] materials scientists have adopted a variety of techniques in the past 10 years to automatically generate phase maps (phase concentrations as a function of composition) from XRD patterns, as recently summarized in a perspective article[4] noting the critical need for improved algorithms that appropriately capture the complexity of materials phase behavior. By developing algorithms at the forefront of artificial intelligence research, we have established a phase mapping tool that models alloying-based shifting of peaks in XRD patterns and incorporates physical constrains based on Gibbs' phase rule. The algorithm performs factor decomposition with an elegant agility that allows users to constrain the spectral demixing based on expert knowledge of the phase behavior, yielding the powerful phase-mapping tool we hereby introduce as AgileFD.

The phase mapping tools developed to date employ a variety of intuitive approaches to the problem of formulating phase maps from a collection of XRD patterns, which generally contain a mixture of phase-pure patterns. Recent efforts have demonstrated that if an expert has sufficient knowledge of the phase map to generate training data, in particular expert-labeled XRD patterns for a variety of phases in the system, the trained machine learning model can map phase behavior.[5] An interesting feature of these supervised machine learning approaches is that phase classification is performed on each XRD pattern sequentially, whereas the remainder of the algorithms discussed below act on the ensemble of XRD patterns to de-mix multi-phase patterns through simultaneous consideration of a wide range of phase mixtures. Clustering XRD patterns based on their similarity[6] is a powerful data reduction tool but since XRD patterns may be clustered due to either a common phase or a common phase mixture (i.e. when a set of composition samples are from the same phase field in the underlying phase diagram), the behavior of pure-clustering techniques is sensitive to the distance metric and requires delicate manual tuning. Initial efforts to address these issues by using constraint programming (CP) to inject physics-based reasoning into clustering have shown promise but have yet to be developed into an effective phase mapping algorithm for experimental data.[7] More recently, by performing clustering in a feature vector space that includes the XRD patterns and sample compositions, compositional clusters can be generated that more closely emulate phase diagrams, and this clustering can be performed quickly enough to perform on-the-fly analysis.[8]

An expert's manual analysis of XRD patterns is most closely emulated with a reasoning-based algorithm in which constituent phases are identified by recognizing sets of peaks that coexist in connected composition regions. By expressing the phase mapping problem as a set of logical requirements within a satisfiability modulo theory (SMT) reasoning framework, we previously demonstrated phenomenal phase mapping performance using pristine, synthetic data.[9] The implementation of this approach for experimental data suffers from practical limitations due to the computational expense for large datasets and the reliance on accurate peak detection in every XRD pattern. As a result, the development of scalable algorithms for (noisy) experimental data has focused on factoring a collection of XRD patterns into a small set of basis patterns (intended to represent individual phases) and the weighting coefficients or "activation" of the basis patterns for each experimental XRD pattern (intended to represent the amount of each phase present in the corresponding material). Since both XRD patterns and phase concentrations are nonnegative, the mathematical description of this approach is non-negative matrix factorization (NMF) in which the XRD patterns for the composition samples are assembled into a matrix $A$ and the phase map solution corresponds to identifying matrices $W$ (whose columns are the basis patterns) and $H$ (which contains the activation of each basis pattern for each sample), such that

$$A \approx W \cdot H. \tag{1}$$

Recent progress in solving the phase mapping problem has thus included implementations of NMF that yield excellent reconstruction of the XRD dataset and produce phase maps that adhere (as closely as possible) to the underlying materials physics.

The various NMF approaches can be succinctly summarized by considering their modelling of 3 principle properties of solid state phase diagrams, particularly for composition libraries synthesized at a fixed temperature: 1. Gibbs' Rule - in thermodynamic equilibrium, the number of coexisting phases in a given material can be no more than the number of components (number of cations in the case of metal oxides) it contains, 2. Connectivity - the set of composition samples containing a given phase must be connected in composition space, 3. Peak Shifting or alloying - the interstitial or substitutional solution of element(s) within a single phase can yield composition-dependent lattice constants that cause XRD patterns to shift as a function of composition. While alloying-based peak shifting can be quite complex for non-cubic phases, in the present work we only consider isotropic lattice expansions that are manifested as uniform shifting of peaks in XRD patterns.

Long et al[10] introduced NMF as a phase mapping strategy, and while this approach is computationally efficient and effective for some datasets, it does not model any of the above properties of phase diagrams, resulting in routine production of non-physical and uninterpretable phase maps. The complement to bare NMF is CombiFD,[11] a powerful factor decomposition approach in which the matrix factorization can be performed under constraints of arbitrary complexity, including the encoding of the above phase diagram properties as hard combinatorial constraints. This approach is analogous to the logic of the SMT algorithm described above and is too computationally intensive for large datasets. GRENDEL[12] is an extension of NMF that was recently adapted for phase mapping by Kusne et al.[13] to enforce the Connectivity constraint by combining NMF with a graphical model of the composition space. GRENDEL remains computationally efficient by iterating between NMF and clustering with an objective function that combines the loss functions of these sub-algorithms, which requires tuning of the relative weights of the loss functions to optimize clustering. Further development of the algorithm is required to enforce Gibbs' Rule, and the reliance on NMF results in an inability to incorporate Peak Shifting. Indeed Peak Shifting has remained a primary challenge in phase mapping, and while powerful approaches for modelling Peak Shifting using dynamic time warping techniques have been proposed,[7, 14] they have yet to be incorporated into scalable phase mapping algorithms.

The AgileFD phase mapping algorithm is based on convolutional non-negative matrix factorization (cNMF),[15] an algorithm that is effective for demixing audio signals, which we have tailored to explicity incorporate Peak Shifting in the matrix factorization process. The key concept is to introduce multiple copies of the basis pattern for each phase, representing multiple alloys of the same phase but with different lattice constants. Gibbs' Rule is also incorporated in AgileFD-Gibbs, an extension of AgileFD that includes constraint optimization of $H$ using Mixed Integer Programming (MIP). AgileFD also enables an expert user to inject knowledge of the phase map through "expert constraints." We have yet to develop an explicit Connectivity constraint since the phase maps we have generated to date do not substantially violate this rule due to the efficacy of our Gibbs' Rule enforcement. As discussed below, our experience has been that these minor Connectivity violations can be easily corrected by incorporating expert feedback into the solutions.

Perhaps the most powerful aspect of AgileFD is its ability to generate both unsupervised and expert-constrained solutions for sizable datasets (i.e. the data matrix $A$ containing ~$10^6$ values corresponding to 100s of XRD patterns each containing 1000s of data points) in a few minutes, enabling the user to interact with solutions via novel data visualizers by altering the number of basis patterns and applying expert constraints until a satisfactory phase map is obtained. The typical analysis of a given XRD dataset proceeds by running AgileFD in unsupervised mode to generate candidate phase mappings that are further tailored with constraints applied by expert analysts.

We demonstrate this process flow for a pseudo-ternary V-Mn-Nb oxide library in which six primary phases are automatically identified with basis patterns that closely match those of known phases. Manual identification of 2 additional, minor phases yields a slightly more refined phase map that is used to interpret high-throughput optical spectroscopy data. While solving the phase behavior of these previously-unexplored compositions demonstrates the power of AgileFD, we additionally highlight its utility in the discovery of functional materials, in particular solar light absorbers.

Using the same V-Mn-Nb oxide library, we constructed a composition map of band gap energy using automated experimentation[16] and Tauc analysis.[17] By correlating band gap energy with the phase concentration and shift parameter from the AgileFD-Gibbs solution, we have discovered new visible-band-gap metal oxides and alloying-based tuning of the direct-allowed band gap energy. Such metal oxide semiconductors are critically missing in the quest to develop efficient solar fuel generators,[18] and the combination of automated phase and band gap mapping described herein will significantly bolster the recent acceleration in the discovery of light absorbers for solar energy applications and beyond.[19]

**Algorithm and Experiments**

*AgileFD algorithm*

In this article, we highlight the innovations of AgileFD, mainly focusing on its incorporation of physical constraints, which drastically differs from previous-reported approaches. For a comprehensive description of the mathematics of AgileFD, we refer the reader to Ref. [20].

Perhaps the most substantial advancement of AgileFD for the phase mapping problem is the efficient modelling of Peak Shifting, which we enable by applying a log-transformation to the XRD patterns. Each XRD pattern can be represented as the scattering intensity ($I$) as a function of $q$, the magnitude of the x-ray scattering vector. The contraction of a crystal lattice due to alloying can be considered as a scaling of the lattice parameter ($a \rightarrow \gamma a$), which corresponds to the inverse scaling of the scattering vector magnitude of each Bragg peak in the XRD pattern ($q_{peak} \rightarrow \gamma^{-1} q_{peak}$). For the present discussion, we consider lattice contraction ($\gamma < 1$) and note below that the method generalizes to lattice contraction and expansion. By performing a log-transformation of $q$, the peak shifting becomes ($\log q_{peak} \rightarrow \log \gamma^{-1} + \log q_{peak}$), and the additive shift can be incorporated in the matrix representation of XRD patterns through a row-shifting operation that we define below. The numerical values of $q$ are not used in matrix factorization, but the $q$-spacing between data points is important when modelling shifting. A consistent model of Peak Shifting is enabled by choosing a constant interval $\rho$ in the $\log q$ space so that the abscissa of the XRD patterns is a geometric series of $q$ values:

$$I_n(\rho^l q_{min}) \text{ where } n \in [0, N-1], l \in [0, L-1], \quad (2)$$

where the $N$ XRD patterns correspond to $N$ composition samples and are represented as the columns of the input data matrix $\mathbf{A}$. By choosing a value of $L$ and using the same range of $q$ values as the source data, these patterns are calculated by resampling the raw XRD patterns, and we typically choose $L$ to be the number of data points in the raw XRD patterns, which is 2082 in the present work.

In traditional NMF, the factorization produces $K$ basis patterns $W_k$, where $K$ is the number of phases used to describe the material system. Our innovative model to account for shifting is enabled by using $K \times M$ patterns in $\mathbf{W}$ that consist of $M$ shifted versions of the basis pattern of each phase. Starting with the "non-shifted" version $W_{k,0}$, the additional $M - 1$ shifted versions are automatically generated in each optimization step by the row-shifting operation:

$$W_{k,m}(\rho^l q_{min}) = \begin{cases} W_{k,0}(\rho^{l-m} q_{min}) & for\ l \in [m, L-1] \\ 0 & for\ l \in [0, m-1] \end{cases}, m \in [1, M-1]. \quad (3)$$

Each shifted version corresponds to a lattice parameter contraction by a factor of $\gamma = \rho^{-m}$, which means that compared to the highest (or lowest) lattice parameter realized in the composition library, lattice contraction (or expansion) by $\rho^{-(M-1)}$ (or $\rho^{(M-1)}$) can be appropriately modelled by activating the corresponding shifted version of the basis pattern. That is, phase $k$ is activated by $N \times M$ coefficients $(H_{n,k,m})$ in the matrix $\boldsymbol{H}$ such that the total activation of phase $k$ for sample $n$ is:

$$h_{n,k} = \sum_{m=0}^{M-1} H_{n,k,m}.$$

(4)

Since multiple terms in this sum can be nonzero, multiple shifted versions of each basis pattern can be used to create a pseudo-continuous model of Peak Shifting as long as the Bragg peaks are relatively wide ($\Delta q_{peak} > \rho \times q_{peak}$), which corresponds to using a sufficiently large value of $L$ such that each peak spans over several data points. The representative shift parameter for a given sample and phase is calculated as a weighted mean:

$$s_{n,k} = \langle \rho^m \rangle_{H_{n,k,m}} = \sum_{m=0}^{M-1} \rho^m H_{n,k,m} \Big/ h_{n,k},$$

(5)

which is the expectation value of $\rho^m$ for each sample and for each basis pattern given the corresponding distribution of activation values.

While this construction of peak shifting is quite intuitive, the pattern shifting cannot be directly incorporated into NMF, which is one of the primary motivations for using cNMF as a starting point in developing AgileFD. Candidate solutions (typically starting with random seeding of $\boldsymbol{H}$ and $\boldsymbol{W}$) are improved in AgileFD through scalable update rules where the loss function is translated into multiplicative matrix operations that enforce gradient descent. In Ref. [20] we derive AgileFD's custom update rules for both $\boldsymbol{H}$ and $\boldsymbol{W}$, which are derived from a loss function based on the generalized Kullback-Leibler divergence, an advancement that is critical for the phase mapping problem and is generally applicable for other source separation problems. With these lightweight update rules, the expansion of both matrices by a factor of $M$ does not substantially prolong convergence time, making this incorporation of Peak Shifting vastly more efficient than the previously-proposed time warping techniques.[7, 14]

While we have developed and continue to develop a variety of update rules to customize the loss function and impose various constraints, we focus on two additional types of constraints in the present work, namely imposing additional expert constraints and enforcing Gibbs' Rule. Expert constraints comprise a variety of methods in which an expert can inject knowledge of the phase behavior, and AgileFD provides convenient mechanisms for encoding constraints through appropriate initialization of $H$ and/or $W$. The multiplicative update rules have the convenient property that if any matrix value is zero in a candidate solution, it will remain zero throughout AgileFD convergence, which makes incorporation of certain constraints as simple as initializing specific matrix elements to zero. For example, to restrict the amount of shifting for a select basis pattern, the user can initialize the $H$ values corresponding to the activation of some of its shifted copies to zero. To restrict the activation of a basis pattern to a select composition region, the user can make its activation zero for all other samples. A novel aspect of AgileFD is that imposing such constraints is substantially more computationally efficient than the (more expressive) incorporation of constraints in combinatorial reasoning algorithms such as CombiFD.[11] The most commonly desired expert constraint for $W$ is to predefine the basis pattern for a phase that is known to exist in the composition library. The corresponding basis pattern can be incorporated in the initialization of $W$, and by not updating that portion of the matrix, the basis pattern is "frozen" in the AgileFD solution.

Gibbs' Rule is incorporated into AgileFD through an extension called AgileFD-Gibbs. In terms of the $h_{n,k}$ values, this rule corresponds to restricting the number of nonzero $h_{n,k}$ to be no greater than the number of components in the composition of sample $n$. The full Gibbs phase rule is actually more restrictive than the one implemented in AgileFD-Gibbs, but we find that restricting the number of coexisting phases according to the number of components is sufficient for practical use. The lack of enforcement of Gibbs' Rule has important consequences on the solution that cannot be ameliorated *ex post facto*. That is, a candidate solution can be "corrected" by zeroing the smallest $h_{n,k}$ values for each $n$ until the solution satisfies Gibbs' Rule, but this guarantees that the corresponding basis patterns are not optimal and that the reconstructed dataset may be "missing" Bragg peaks that are in the source data. More generally, this simple "correction" algorithm may not generate optimal solutions even after additional optimization of the basis patterns. To more elegantly and optimally enforce Gibbs' Rule, we combine AgileFD with a MIP algorithm that modifies a candidate solution. While $W$ is fixed, the MIP algorithm generates a new activation matrix by finding the optimal $K \times M$ activation coefficients for each sample that adhere to Gibbs Rule, and the resulting candidate solution is then further optimized using AgileFD. The MIP algorithm is applied to each sample *independently*, and the execution of $N$ small MIP routines does not substantially prolong the AgileFD-Gibbs solution time. It is worth noting the similarity between the utility of the MIP algorithm and the expert constraints noted above; the MIP-modified solution assigns the appropriate number of values in $H$ to zero such that subsequent candidate solutions adhere to Gibbs' Rule.

With an AgileFD or AgileFD-Gibbs solution in hand, the collection of basis patterns $W_{k,0}$ and the composition maps of $h_{n,k}$ and $s_{n,k}$ provide an intuitive visualization of the phase behavior. Further quantification of phase concentrations and lattice parameters require identification of the crystal structure for each basis pattern, which was performed in the present work through search and match with the entries in the International Center for Diffraction Data (ICDD) database using DIFFRAC.SUITE™ EVA software. Using relative intensity ratios and tabulated Bragg peaks from the ICDD entries (see SI for details), the relative total scattering intensity per mole of metal ($\vartheta_k$) was calculated for each phase, enabling the relative molar activation of each phase in each sample to be calculated as

$$^{rel}P_{n,k} = \frac{h_{n,k}}{\vartheta_k} \int W_{k,0},$$

(6)

where the integral over the basis pattern is analogous to the integration over the ICDD pattern. The $k$ phase fractions for each sample are then provided through normalization of these relative molar activations:

$$P_{n,k} = {}^{rel}P_{n,k} \Big/ \sum_{k=0}^{K-1} {}^{rel}P_{n,k},$$

(7)

which is the solution to the phase map problem under the approximation that the total relative scattering factor of each phase matches that of the ICDD pattern and does not vary substantially with alloy composition within the phase.

To provide a more intuitive visualization of the weighted shift parameters, $s_{n,k}$, they are converted to lattice expansion parameters $\gamma_{n,k} = s_{n,k}^{-1}$. By determining the lattice expansion parameter $\gamma_{k,ICDD}$ that corresponds to the shift of $W_{k,0}$ that best matches the respective ICDD pattern, the relative shift of each phase in each sample compared to its ICDD entry is the ratio of $\gamma_{n,k}$ to $\gamma_{k,ICDD}$.

*Library Synthesis*

The continuous composition spread of V, Mn, and Nb was synthesized by reactive magnetron co-sputtering in the presence of $O_2$ and Ar gas using elemental sources arranged symmetrically with respect to the 100 mm Si/SiO$_2$ substrate. The deposition proceeded for about 10 hours with the RF power of the V, Mn, Nb source fixed at 150 W, 115 W, and 81 W, respectively. The spatial variation in deposition rate from each source resulted in a continuous thin film with composition gradient of the order of 0.5 at.% mm$^{-1}$. The variation in the deposition rates amongst different sources resulted in thickness variation across the library within a factor of two of the center thickness of 400 nm. The as-deposited oxide composition library was oxidized in air at

883 K for 1 hour, producing the (V-Mn-Nb)O$_x$ library for which oxygen composition is unknown and samples are represented by their cation composition.

*Composition and Structure Measurements and Analysis*

XRD data was acquired using a custom HiTp setup incorporated into the bending-magnet beamline 1-5 of the Stanford Synchrotron Radiation Light source (SSRL) at SLAC National Accelerator Laboratory. A detailed description of the experiment is provided by Gregoire et al.,[2a] and the characterization of the (V-Mn-Nb)O$_x$ library employed a monochromated 13 keV source in reflection scattering geometry with an XRD image detector (Princeton Quad-RO 4320). Using a 1 mm$^2$ footprint, $N = 317$ samples were acquired on a square grid with 4.5 mm pitch and within a radius of 45 mm from the center of the 100 mm-diameter library. Diffraction images were processed into XRD patterns, $I_n(q)$, using WxDiff software[21] and further processed with a custom background subtraction algorithm using cubic splines.

The XRF measurements were performed on an EDAX Orbis Micro-XRF system (EDAX Inc., Mahwah, NJ) with an x-ray beam approximately 1 mm in diameter. The V K, Mn K and Nb K XRF intensities were extracted from the Orbis software and converted to normalized V-Mn-Nb compositions using relative sensitivity factors calibrated at substrate center via a separate composition measurement. The calibration composition was measured on an Oxford Instruments X-Max 80 mm$^2$ energy dispersive x-ray spectroscopy (EDS; Oxford Instrument, Concord, MA) detector on a FEI Nova NanoSEM 450 (FEI, Hillsboro, OR). The absolute composition uncertainty for this EDS measurement is 10 at.%, and the XRF-determined relative compositions, which enable composition-property maps, have approximately 1 at.% resolution.[4,22]

*Optical Characterization*

Optical characterization of the (V-Mn-Nb)O$_x$ library was performed on a custom HiTp diffuse reflectance (DR) instrument built to analyze light absorber thin films on opaque substrates. The computer-automated experiment employed a 200 W Hg(Xe) lamp (Newport/Oriel Apex) and an integrating sphere (Ocean Optics ISP-50-8-R-GT) fiber-coupled to a spectrometer (Spectral Products, Inc. SM303), with further details provided by Mitrovic at et al.[16] DR spectra were acquired on a square grid of 1521 positions with a pitch of 2.032 mm, spot size of approximately 1 mm, and integration time of 0.025 s per spectrum, and typically three spectra were averaged to produce the DR spectrum for each sample. The absorption coefficient (α) was calculated up to a factor of a spectral scattering factor, which we approximate to be energy-independent, using the Kubelka-Munk radiative transfer model[23] which enabled calculation of the normalized Tauc property:

$$TP = (\alpha h\nu)^{1/n}/\max((\alpha h\nu)^{1/n}), \qquad (8)$$

where *n* is the Tauc exponent whose value is ½ for analysis of direct-allowed (DA) electronic transitions. The plots of DA Tauc property ($TP^{DA}$) vs photon energy ($h\nu$) were not only automatically generated but also automatically interpreted using a recently-developed algorithm[17b] that mimics expert judgement. The algorithm either estimates the DA band gap energy or determines that the band gap cannot be confidently estimated from the Tauc plot, which can occur when the band gap energy is beyond the range of the spectrometer, the sample is not sufficiently absorbing, or the direct-allowed band gap signature is convoluted by the presence of multiple phases with comparable contributions to the DR signal. We note that for discovery of photoabsorbers, indirect-allowed (IA) transitions are of primary interest only when the IA band gap is substantially lower than DA band gap. For the DA results shown here, no IA band gaps were identified at energies more than 0.3 eV below the direct-allowed gap.

The optical and XRF measurements were performed on the same set of library samples. The XRD measurements were performed on a coarser grid of library positions, and the composition for each XRD sample was calculated using linear interpolation in the Cartesian library position space. The results of XRD analysis ($\gamma_{n,k}$ and $P_{n,k}$) were interpolated to the optical and XRF samples through linear interpolation in the ternary composition space. The composition spread library and compositions of the samples characterized are shown in Figure 1.

**Results and Discussion**

The phase behavior of the (V-Mn-Nb)O$_x$ library was analyzed using AgileFD-Gibbs in unsupervised mode and then with additional expert constraints. Manual visualization of the XRD patterns revealed substantial Peak Shifting, prompting our use of $M = 10$ shifted copies of each basis pattern, which corresponds to a maximum lattice expansion of 1.0072 with respect to the basis pattern with the lowest lattice constant. Several values of $K$ were attempted, revealing that the $K = 6$ solution appeared physically sound and that higher values of $K$ neither substantially improved the data reconstruction nor produced basis patterns that were distinctive from those of the $K = 6$ solution. A more objective determination of appropriate values for parameters such as $K$ and $M$ is the subject of ongoing research and is beyond the scope of the present work.

Using the basis patterns from the $K = 6$ solution, we performed phase matching and identified the first 6 phases in Table 1. That is, upon fixing the values of $K$ and $M$, the unsupervised AgileFD-Gibbs algorithm produced 6 basis patterns that match those of known phases despite substantial overlap of the basis patterns, which is a testament to the excellent source separation enabled by the update rules and Gibbs rule constraint. To demonstrate how the identification of 6 phases was enabled by the Peak Shifting feature of AgileFD, we show the AgileFD $M = 1$ and AgileFD-Gibbs $M = 10$ solutions in Figure 2. With $M = 1$, the dimension of the activation matrix $H$ is equivalent to that of NMF, motivating our labelling of this solutions as "~NMF", but the algorithms are not equivalent due to both the log transformation and gradient descent with custom update rules used in AgileFD. While some basis patterns and their phase distributions are similar between the AgileFD-Gibbs and ~NMF solutions, the lack of Peak Shift modelling has an important consequence in the ~NMF solution. Inspection of the $k = 5$ basis pattern in Figure

2 reveals that while the AgileFD-Gibbs solution is an excellent match to the ICDD pattern, the ~NMF pattern is a very poor match. Instead of separating this ICDD phase, the data is better reconstructed in the ~NMF solution by using 2 basis patterns ($k = 2$ and $k = 5$) to model the $k = 2$ solution of AgileFD-Gibbs. The $k = 2$ and $k = 5$ basis patterns in the ~NMF solution are essentially shifted versions of each other and activated in complementary portions of the composition space. The ~NMF solution fails to identify the $k = 5$ phase from the AgileFD-Gibbs solution as a direct consequence of the inability to model Peak Shifting.

Additional subtle but important difference exist between the ~NMF and AgileFD-Gibbs solution, and perhaps the illustrative example for the present discussion is that the AgileFD-Gibbs solution adheres to the Connectivity requirement much more closely than the ~NMF solution even though this constraint was not imposed. The enforcement of the Gibbs rule has important implications for optimizing not only the phase distributions but also the basis patterns. Figure 2 also shows that the AgileFD-Gibbs solutions provides the lattice constant for each phase, which for several phases varies systematically through the composition space, a strong indicator of alloying.

Upon careful inspection of the AgileFD-Gibbs solution we also noticed some opportunities to further improve the phase map via expert constraints in AgileFD-Gibbs. We proceeded to manually inspect the compositions closest to the end-member binary oxides. For Mn oxide the AgileFD-Gibbs solution produced 2 different Mn oxide phases ($k = 0$ and $k = 5$). We have commonly observed the coexistence of these phases in Mn-containing libraries of metal oxides due to the lack of a strong thermodynamic differentiation under the library synthesis conditions. Since the library compositions contain substantial amounts of Nb and V that may alloy into these phases, the alloyed metal oxides may indeed coexist in thermodynamic equilibrium in this ternary composition space. Through manual inspection of XRD patterns containing these phases, we confirmed that both phases are present and in fact they often coexist. It is worth noting that without prior knowledge of the XRD patterns of pure phases, any source separation algorithm cannot robustly produce phase-pure basis patterns for phases that always coexists. For the Mn oxides, AgileFD-Gibbs provides sufficient separation of the phases that they were identified in ICDD search and match, and we enforced complete separation by seeding $W$ with simulated basis patterns obtained by convoluting the ICDD patterns with a Gaussian peak shape ($\sigma = 0.13$ nm$^{-1}$). Inspection of the most Nb-rich samples revealed that, as suggested by Figure 2, the library samples are far enough from the Nb end-member composition that no Nb oxides are observed. Inspection of the V-rich composition revealed the presence of 2 minor phases, $V_2O_5$ and $NbVO_5$, whose relatively weak intensity and existence in only a small fraction of the composition samples gives their signals little contribution to the system-wide loss function. While AgileFD-Gibbs may factor these phases with an extended dataset that includes compositions closer to the stoichiometries of these phases, we amended this shortcoming of the source data by introducing and freezing a new basis pattern for both $V_2O_5$ and $NbVO_5$. Since these phases were only observed in V-rich compositions, we additionally seeded the corresponding values of $H$ with zeros for all samples with V concentration less than 0.45. The expanded $K = 8$ solution with expert constraints on four phases was generated using AgileFD-Gibbs, producing the solution shown in Figure 3.

The addition of 2 phases and expert constraints enabled more accurate modelling of every phase, leading to slight changes (compared to the AgileFD-Gibbs solution of Fig. 2) in the basis patterns and phase maps even for the phases that were not constrained. The agreement between $k = 1,2,4$ and the respective basis patterns is remarkable, and the agreement for $k = 3$ is quite good but not perfect. Upon manual inspection of samples with high activation of this phase, we believe this basis pattern accurately models the complex phase behavior in this composition region. It appears that the $Mn_3V_2O_8$ coexists with minor fractions of 2 polytypes of $Mn_2V_2O_7$. We previously reported[24] on the coexistence of these polytypes in thin film (V-Mn)O$_x$ libraries and did not pursue further disambiguation in the phase map solution since all 3 of these manganese vanadates only appear as minor phases since the composition samples are all beyond ~10 at.% of the V-Mn binary line.

With the phase map of Fig. 3 in hand, we turn to the optical characterization data to investigate if the composition library contains any metal oxide phases of interest for solar light absorption. While the automated Tauc analysis algorithm did not identify a DA transition for some samples, DA band gap energies ($E_g^{DA}$) were estimated for 1329 samples and are mapped in composition space in Figure 4. While comparisons of the band gap data and phase map reveal a number of interesting structure-property relationships, for the present purposes we provide a detailed interpretation of $MnV_2O_6$, corresponding to $k = 4$ in Figure 3. It is worth noting that for mixed-phase samples, the majority phase does not necessarily provide the majority contribution to the Tauc plot due to possible differences in the optical absorption strength and band gap energies of different phases. To infer which composition samples exhibit a band gap representative of $MnV_2O_6$, we show the band gap energy as function of $MnV_2O_6$ phase fraction in Figure 5, in which the sample points are colored according to the relative lattice constant from the $k = 4$ phase map of Figure 3. With low phase fraction, the band gap energy is most likely due to a different phase, so inspecting the convergence of the band gap energy as the phase fraction increases provides an indication of the phase fraction at which $MnV_2O_6$ is providing the major contribution to the band gap measurement. Figure 5 reveals that samples with phase fraction in excess of 0.8 exhibit a band gap in the 1.8-2.08 range, and that the variation within this range is correlated with the lattice parameter.

To further investigate this relationship between band gap and lattice parameter in $MnV_2O_6$, we extract the samples with phase purity in excess of 0.8 and visualize in Figure 6 both the original XRD patterns and the relative lattice parameter from AgileFD-Gibbs. This visualization provides a more detailed understanding of the phase behavior in this composition region. At the lowest V concentration (composition $V_{0.19}Mn_{0.38}Nb_{0.43}O_x$), the lattice parameter is approximately 1.009 times larger than the ICDD entry and as the V concentration is increased to 0.5, the relative lattice constant systematically lowers and trends toward 1 as the cation composition gets closer to the $Mn_{0.33}V_{0.67}$ composition of stoichiometric $MnV_2O_6$. The composition library does not extend to this composition, and in the composition region noted in Figure 6, there is a phase boundary at V concentration of 0.5 where the alloying of $MnV_2O_6$ does not continue into more V-rich compositions and this phase becomes mixed with $NbVO_5$. The highlighted composition region in Figure 6 contains phase-pure $MnV_2O_6$ (within the detectability limit of other phases) and exhibits alloying-based peak shifting with respect to both Mn and Nb, suggesting that the Mn concentration varies by ±4 at.% from the phase stoichiometry and that Nb is soluble up to approximately 25 at.%. For this non-cubic crystal structure, the lattice constant determined by AgileFD-Gibbs is an "average" lattice constant but not necessarily the mean of the three lattice parameters. Figure 6a shows that most peaks shift uniformly with some indication of non-isotropic lattice expansion that we do not model in the present work and may provide insights into the elemental site substitutions of this oxide alloy.

To investigate the impact of the alloying-induced lattice parameter shift on band gap energy, we consider the samples in the composition region highlighted in Figure 6. As shown in Figure 7, the band gap energy of these samples spans a nearly 0.2 eV range and systematically increases with increasing relative lattice parameter. We note that while the band gaps of these oxide alloys have not been previously reported, the trend in Figure 7 is commensurate with reported direct-allowed band gap value of 1.8-1.95 eV for $MnV_2O_6$ (where the relative lattice parameter is 1).[25] This alloying-induced variation in the band gap energy is well-studied in semiconductors based on main-group elements but is rarely observed in metal oxides, particularly in the 1.8-2.1 eV range that is of primary interest for photoanodes in solar hydrogen generators and other applications.[18] While substantial additional research is required to fully understand the alloying and band gap tuning in V-Mn-Nb oxides, the results presented in Figures 2-7 highlight the excellent performance of the AgileFD suite of algorithms and their enabling ability to conduct materials science research in high order composition spaces.

## Summary

We introduce AgileFD, the first scalable phase mapping algorithm for combinatorial XRD data that models alloying-based peak shifting and imposes Gibbs' phase rule. The importance of encoding these properties of solid state phase diagrams into a source separation algorithm is discussed using the $(V-Mn-Nb)O_x$ system where unsupervised mapping of six phases is directly enabled by these capabilities of AgileFD. Several of the 317 XRD patterns from the composition library contain small signals from minor phases that are not amenable to unsupervised factorization but can be modelled by AgileFD through straightforward injection of expert knowledge. The resulting 8-phase solution reveals alloying in ternary oxide phases such as $MnV_2O_6$. By combining the AgileFD solution with band gap energies obtained from automated Tauc analysis of high throughput spectroscopy data, we identify band gap tuning of nearly 0.2 eV as a function of lattice parameter and V composition in the energy range of interest for solar applications. The identification of this family of promising solar light absorbers was enabled through a multi-disciplinary effort in which materials-motivated advancements of computer science techniques produced powerful new algorithms for accelerating materials discovery.

## Acknowledgements

The experimental work was performed in the Joint Center for Artificial Photosynthesis (JCAP), a DOE Energy Innovation Hub supported through the Office of Science of the U.S. Department of Energy under Award No. DE-SC0004993. The algorithm development is supported by NSF

awards CCF-1522054 and CNS-0832782 (Expeditions), CNS-1059284 (Infrastructure), and IIS-1344201 (INSPIRE); and ARO award W911-NF-14-1-0498. Use of the Stanford Synchrotron Radiation Lightsource, SLAC National Accelerator Laboratory, is supported by the U.S. Department of Energy, Office of Science, Office of Basic Energy Sciences under Contract No. DE-AC02-76SF00515. The authors thank Apurva Mehta and Douglas G. Van Campen for assistance with collection of synchrotron XRD data.

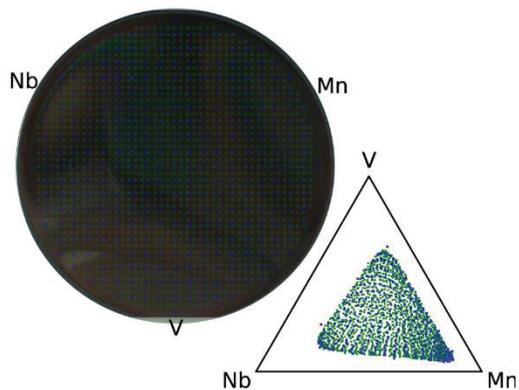

Figure 1. A white light image of the (V-Mn-Nb)O$_x$ composition library on 100 mm Si/SiO$_2$ substrate is shown with elemental labels indicating the orientation of the sputter deposition sources. The grid of library locations for XRD (blue) and both XRF and optical characterization (green) are shown along with an additional plot showing the XRF-determined compositions of these points.

Table 1: The primary phases in the AgileFD-Gibbs solutions for $K=6$ and $K=8$ are listed along with their ICDD entries, which are overlaid in the basis pattern plots of Figures 2 and 3. The $\vartheta_k$ values used to calculate phase fractions are also shown.

| k, phase index | Formula Unit (crystal system) | ICDD entry number | Relative total scattering intensity per mole of metal ($\vartheta_k$) |
| --- | --- | --- | --- |
| 0 | $Mn_2O_3$ (cubic) | 01-071-0636 | 923.2 |
| 1 | $V_{2.38}Nb_{10.7}O_{32.7}$ (orthorhombic) | 01-079-8393 | 2195.7 |
| 2 | $MnNb_2O_6$ (orthorhombic) | 01-072-0484 | 2195.2 |
| 3 | $Mn_3V_2O_8$ (unknown) | 00-039-0091 | 739.2[26] |
| 4 | $MnV_2O_6$ (monoclinic) | 01-072-1837 | 911.8 |
| 5 | $Mn_3O_4$ (tetragonal) | 01-080-0382 | 964.5 |
| 6 | $NbVO_5$ (orthorhombic) | 00-046-0046 | 1706.2 |
| 7 | $V_2O_5$ (orthorhombic) | 00-041-1426 | 678.0[27] |

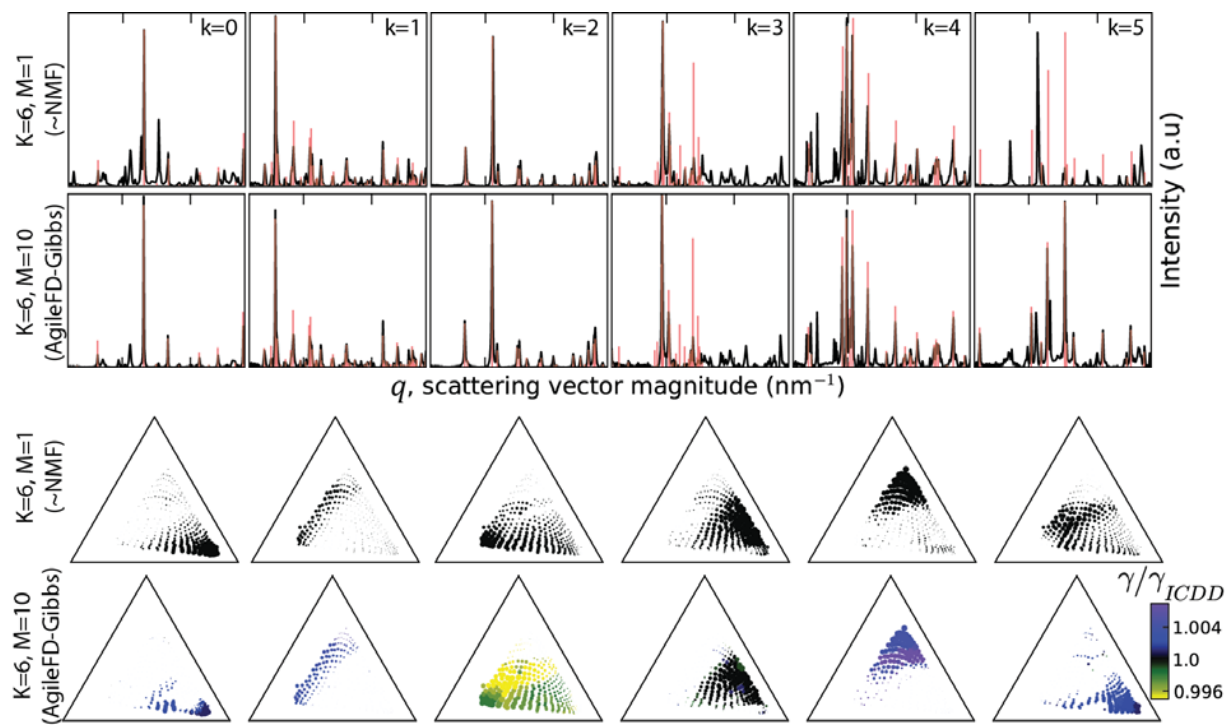

Figure 2. Six-phase solutions for the (V-Mn-Nb)O$_x$ library with algorithm parameters noted in the left-hand labels. The "~NMF" solution is the AgileFD solution with the Peak Shifting removed by setting $M = 1$. The basis patterns are plotted along with the ICDD patterns listed in Table 1. The map of each phase is shown as a composition plot where the point size represents the phase fraction $P_{n,k}$ and the color represents the relative lattice constant compared to the respective basis pattern, which is aligned to the best-match with the ICDD pattern.

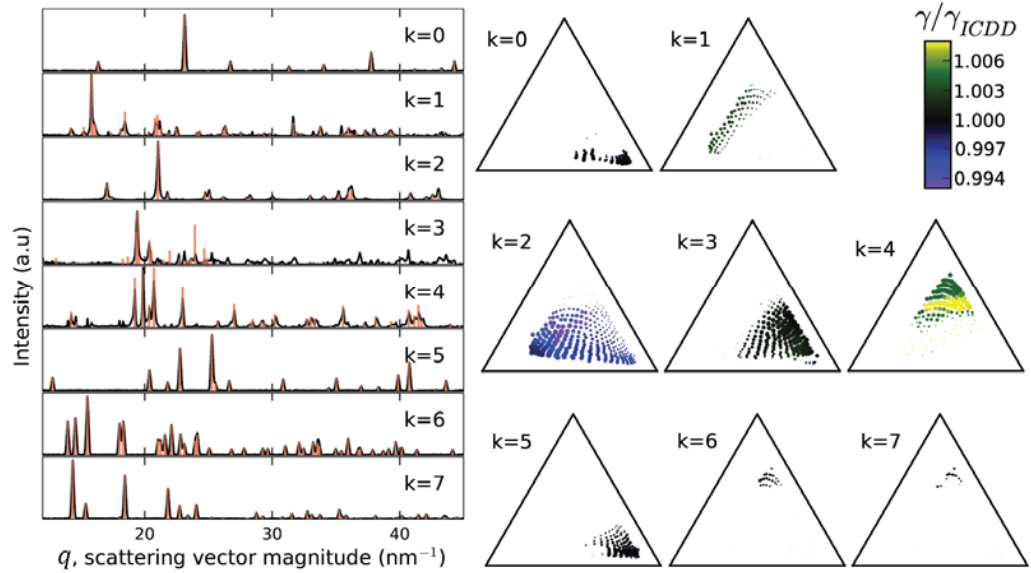

Figure 3. The AgileFD-Gibbs solution with $K = 8$, $M = 10$ and expert constraints applied to $k = 0,5,6,7$. The map of each phase is shown as a composition plot where the point size represents the phase fraction $P_{n,k}$ and the color represents the relative lattice constant, as described in Figure 2.

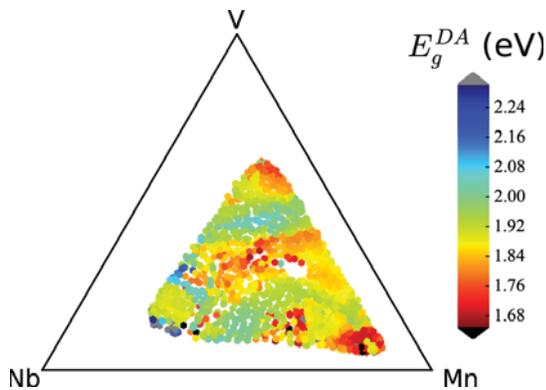

Figure 4. The composition map of the DA band gap energy from automated Tauc analysis is shown for 1329 composition samples.

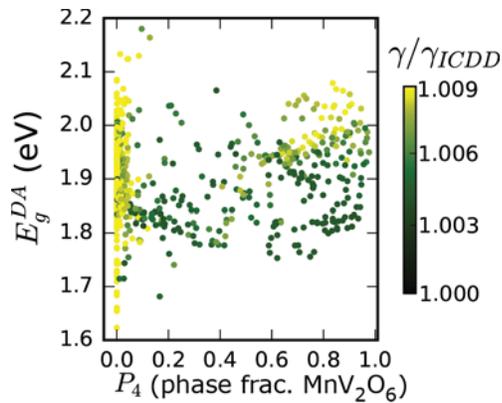

figure 5. The DA band gap energy is plotted against the $k = 4$ phase fraction from Figure 3, demonstrating that when the phase fraction of $MnV_2O_6$ is above 0.8, the band gap value systematically varies with the relative lattice parameter.

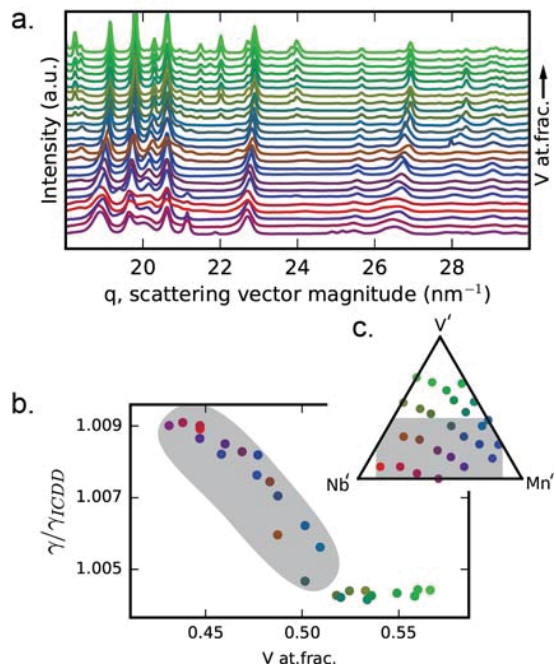

Figure 6. a.) For the 26 samples with phase fraction of $MnV_2O_6$ in excess of 0.8, the series of XRD patterns (in the $q$-range with primary ICDD peaks) are arranged according to the V concentration. The top 10 patterns show a small amount of $NbVO_5$, and the bottom 16 patterns exhibit systematic shifting of the $MnV_2O_6$ peaks with respect to V concentration. b.) The relative lattice constant for the 26 samples is shown and the 16 samples with no $NbVO_5$ are indicated by a gray region. c.) The 26 samples are shown in a composition plot with end-members $V'=V_{0.61}Mn_{0.29}Nb_{0.09}O_x$, $Mn'=V_{0.43}Mn_{0.47}Nb_{0.1}O_x$, and $Nb'=V_{0.43}Mn_{0.29}Nb_{0.28}O_x$. The composition gray composition region contains the same 16 samples as that in b. The data for each sample is colored according to its composition in all 3 plots.

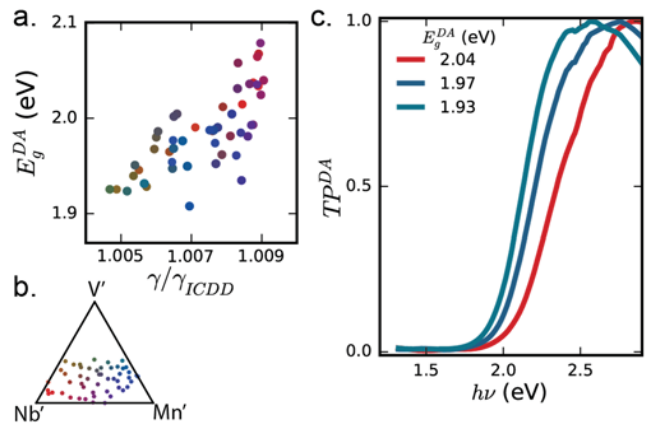

Figure 7: a.) The band gap values from the composition region containing high phase purity $MnV_2O_6$ (see Figure 6) are plotted against the relative lattice parameter, demonstrating alloying-based tuning of the band gap energy. b.) The sample compositions using the same composition-color scale revealing that among these samples, the highest band gap energy and lattice parameters are found with the most Nb-rich compositions. c.) Representative Tauc plots for 3 samples with line colors matching the samples' colors in b. The band gap energies produced by the automated Tauc algorithm are listed in the legend.

**References**

1.	Green, M. L.; Takeuchi, I.; Hattrick-Simpers, J. R., Applications of high throughput (combinatorial) methodologies to electronic, magnetic, optical, and energy-related materials. *J. Appl. Phys.* **2013,** *113* (23), 231101-231101.
2.	(a) Gregoire, J. M.; Van Campen, D. G.; Miller, C. E.; Jones, R.; Suram, S. K.; A., M., High Throughput Synchrotron X-ray Diffraction for Combinatorial Phase Mapping. *J. Synchr. Rad.* **2014,** *21* (6), 1262-1268; (b) Gregoire, J. M.; Dale, D.; Kazimirov, A.; DiSalvo, F. J.; van Dover, R. B., High energy x-ray diffraction/x-ray fluorescence spectroscopy for high-throughput analysis of composition spread thin films. *Rev. Sci. Instrum.* **2009,** *80* (12), 123905-123905.
3.	Barr, G.; Dong, W.; Gilmore, C. J., PolySNAP: a computer program for analysing high-throughput powder diffraction data. *Journal of Applied Crystallography* **2004,** *37* (4), 658-664.
4.	Hattrick-Simpers, J. R.; Gregoire, J. M.; Kusne, A. G., Perspective: Composition–structure–property mapping in high-throughput experiments: Turning data into knowledge. *APL Mater.* **2016,** *4* (5), 053211.
5.	(a) Bunn, J. K.; Han, S.; Zhang, Y.; Tong, Y.; Hu, J.; Hattrick-Simpers, J. R., Generalized machine learning technique for automatic phase attribution in time variant high-throughput experimental studies. *Journal of Materials Research* **2015,** *30* (7), 879-889; (b) Bunn, J. K.; Hu, J.; Hattrick-Simpers, J. R., Semi-Supervised Approach to Phase Identification from Combinatorial Sample Diffraction Patterns. *JOM* **2016,** *68* (8), 2116-2125.
6.	Long, C. J.; Hattrick-simpers, J.; Murakami, M.; Srivastava, R. C.; Takeuchi, I.; Karen, V. L.; Li, X., Rapid structural mapping of ternary metallic alloy systems using the combinatorial approach and cluster analysis. *Rev. Sci. Instrum* **2007,** *78* (7).

7.	Lebras, R.; Damoulas, T.; Gregoire, J. M.; Sabharwal, A.; Gomes, C. P.; van Dover, R. B., Constraint Reasoning and Kernel Clustering for Pattern Decomposition With Scaling. *Proc. 17th Intl. Conf. on Principles and Practice of Constraint Programming* **2011,** *6876*, 508-522.
8.	Kusne, A. G.; Gao, T.; Mehta, A.; Ke, L.; Nguyen, M. C.; Ho, K.-M.; Antropov, V.; Wang, C.-Z.; Kramer, M. J.; Long, C.; Takeuchi, I., On-the-fly machine-learning for high-throughput experiments: search for rare-earth-free permanent magnets. *Sci. Rep.* **2014,** *4*, 6367.
9.	Ermon, S.; Le Bras, R.; Gomes, C. P.; Selman, B.; van Dover, R. B., SMT-Aided Combinatorial Materials Discovery. In *Proceedings of the 15th International Conference on Theory and Applications of Satisfiability Testing (SAT)*, 2012; pp 172-185.
10.	Long, C. J.; Bunker, D.; Li, X.; Karen, V. L.; Takeuchi, I., Rapid identification of structural phases in combinatorial thin-film libraries using x-ray diffraction and non-negative matrix factorization. *Rev. Sci. Instrum.* **2009,** *80* (10), 103902-103902.
11.	Ermon, S.; Le Bras, R.; Suram, S. K.; Gregoire, J. M.; Gomes, C. P.; Selman, B.; Van Dover, R. B., Pattern Decomposition with Complex Combinatorial Constraints: Application to Materials Discovery. In *Proceedings of the Twenty-Ninth Conference on Artificial Intelligence (AAAI-15)*, North America, 2015.
12.	Massoudifar, P.; Rangarajan, A.; Zare, A.; Gader, P. In *An integrated graph cuts segmentation and piece-wise convex unmixing approach for hyperspectral imaging*, IEEE GRSS Workshop Hyperspectral Image Signal Processing: Evolution Remote Sensing (WHISPERS), 2014.
13.	Kusne, A. G.; Keller, D.; Anderson, A.; Zaban, A.; Takeuchi, I., High-throughput determination of structural phase diagram and constituent phases using GRENDEL. *Nanotechnology* **2015,** *26* (44), 444002.
14.	Baumes, L. A.; Moliner, M.; Nicoloyannis, N.; Corma, A., A reliable methodology for high throughput identification of a mixture of crystallographic phases from powder X-ray diffraction data. *Cryst. Eng. Comm.* **2008,** *10* (10), 1321-1321.
15.	Smaragdis, P. In *Non-negative matrix factor deconvolution; extraction of multiple sound sources from monophonic inputs*, International Conference on Independent Component Analysis and Signal Separation, Springer: 2004; pp 494-499.
16.	Mitrovic, S.; Cornell, E. W.; Marcin, M. R.; Jones, R. J.; Newhouse, P. F.; Suram, S. K.; Jin, J.; Gregoire, J. M., High-throughput on-the-fly scanning ultraviolet-visible dual-sphere spectrometer. *Rev. Sci. Instrum.* **2015,** *86* (1), 013904.
17.	(a) Suram, S. K.; Newhouse, P. F.; Zhou, L.; Van Campen, D. G.; Mehta, A.; Gregoire, J. M., High Throughput Light Absorber Discovery, Part 2: Establishing Structure-Band Gap Energy Relationships. *ACS Combinatorial Science* **2016**; (b) Suram, S. K.; Newhouse, P. F.; Gregoire, J. M., High Throughput Light Absorber Discovery, Part 1: An Algorithm for Automated Tauc Analysis. *ACS Combinatorial Science* **2016**.
18.	Hu, S.; Xiang, C.; Haussener, S.; Berger, A. D.; Lewis, N. S., An analysis of the optimal band gaps of light absorbers in integrated tandem photoelectrochemical water-splitting systems. *Energy Environ. Sci.* **2013,** *6* (10), 2984-2984.
19.	(a) Zhou, L.; Yan, Q.; Shinde, A.; Guevarra, D.; Newhouse, P. F.; Becerra-Stasiewicz, N.; Chatman, S. M.; Haber, J. A.; Neaton, J. B.; Gregoire, J. M., High Throughput Discovery of Solar Fuels Photoanodes in the CuO–V2O5 System. *Adv. Energy Mater.* **2015,** *5*, 1500968; (b) Thienhaus, S.; Naujoks, D.; Pfetzing-Micklich, J.; Konig, D.; Ludwig, A., Rapid identification of areas of interest in thin film materials libraries by combining electrical, optical, X-ray diffraction, and mechanical high-throughput measurements: a case study for the system Ni-Al. *ACS combinatorial science* **2014,** *16* (12), 686-94; (c) Subramaniyan, A.; Perkins, J. D.; O'Hayre, R. P.; Ginley, D. S.; Lany, S.; Zakutayev, A., Non-equilibrium synthesis, structure, and opto-electronic properties of Cu2−2x Zn x O alloys. *Journal of Materials Science* **2015,** *50* (3), 1350-1357.

20.     Xue, Y.; Bai, J.; Le Bras, R.; Bernstein, R.; Bjorck, J.; Longpre, L.; Suram, S. K.; Suram, S. K.; Gregoire, J.; Gomes, C. P., Phase-Mapper: An AI Platform to Accelerate High Throughput Materials Discovery. *ArXiv e-prints* **2016**.
21.     Mannsfeld, S. C.; Tang, M. L.; Bao, Z., Thin film structure of triisopropylsilylethynyl-functionalized pentacene and tetraceno[2,3-b]thiophene from grazing incidence X-ray diffraction. *Adv. Mater.* **2011,** *23* (1), 127-31.
22.     Le Bras, R.; Bernstein, R.; Gregoire, J. M.; Suram, S. K.; Gomes, C. P.; Selman, B.; van Dover, R. B., A Computational Challenge Problem in Materials Discovery: Synthetic Problem Generator and Real-World Datasets. In *Proceedings of the Twenty-Eighth Conference on Artificial Intelligence (AAAI-14)*, Québec City, Canada, 2014.
23.     Kubelka, P., New Contributions to the Optics of Intensely Light-Scattering Materials. Part I. *J. Opt. Soc. Am.* **1948,** *38* (5), 448-457.
24.     Yan, Q.; Li, G.; Newhouse, P. F.; Yu, J.; Persson, K. A.; Gregoire, J. M.; Neaton, J. B., Mn2V2O7: An Earth Abundant Light Absorber for Solar Water Splitting. *Adv En Mater* **2015,** *5* (8), 1401840.
25.     Zhang, W.; Shi, L.; Tang, K.; Liu, Z., Synthesis, surface group modification of 3D MnV2O6 nanostructures and adsorption effect on Rhodamine B. *Materials Research Bulletin* **2012,** *47* (7), 1725-1733.
26.     This ICDD entry does not have relative scattering information. The average value from Mn2V2O7 polytypes (01-073-6361,  01-089-0483) is used as a proxy.
27.     This ICDD entry does not have relative scattering information. The average value from V2O5 (00-041-1426) and Nb2O5 (04-007-2424) is used as a proxy.